\documentclass[aps,prl,twocolumn,superscriptaddress]{revtex4}

\usepackage[dvips]{graphicx}
\usepackage{latexsym}
\usepackage{graphicx}
\usepackage{epsfig}
\usepackage[english]{babel}
\usepackage[latin1] {inputenc}
\usepackage{amsmath,amsfonts,amssymb}
\usepackage{bm}

\newcommand \be {\begin{equation}}
\newcommand \bea {\begin{eqnarray}}
\newcommand \ee {\end{equation}}
\newcommand \eea {\end{eqnarray}}
\newcommand \bed {\begin{displaymath}}
\newcommand \eed {\end{displaymath}}

\newcommand{\bit}{\begin{itemize}}
\newcommand{\eit}{\end{itemize}}

\newcommand{\bgamma}{\mbox{\boldmath{$\gamma$}}}
\newcommand{\ba}{\mbox{\boldmath{$a$}}}

\newcommand{\bomega}{\mbox{\boldmath{$\omega$}}}

\newcommand{\bW}{\mbox{\boldmath{$W$}}}

\newcommand{\bS}{\mbox{\boldmath{$S$}}}

\newcommand{\bgar}{\begin{eqnarray}}
\newcommand{\enar}{\end{eqnarray}}

\begin{document}

\title{Opinion dynamics and decision of vote in bipolar political systems}
\author{Filippo Caruso$^{3}$ and Paolo Castorina$^{2,}$}
\affiliation{Dept. of Physics, University of Catania
\quad and \quad
$^{2}$INFN, Sezione di Catania,\\
Via S. Sofia 64, I-95123, Catania, Italy\\
$^{3}$Scuola Superiore di Catania, Via S. Paolo 73, 95123 Catania,
Italy \\filippo.caruso@ct.infn.it, paolo.castorina@ct.infn.it}

\date{\today}

\begin{abstract}
A model of the opinion dynamics underlying the political decision
is proposed. The analysis is restricted to a bipolar scheme with a
possible third political area. The interaction among voters is
local but the final decision strongly depends on global effects
such as, for example, the rating of the governments. As in the
realistic case, the individual decision making process is
determined by the most relevant personal interests and problems.
The phenomenological analysis of the national vote in Italy and
Germany has been carried out and a prediction of the next Italian
vote as a function of the government rating is presented.

Keywords: disordered systems, opinion dynamics
\end{abstract}

\maketitle

\section{Introduction}

The study of complex systems entered social science in order to
understand how self-organization, cooperative effects and
adaptation arise in social communities. In this context the use of
simple automata or dynamical models often elucidates the
underlying dynamics of the observed behaviour (1-13).

In particular some models have been proposed to describe the
opinion dynamics and the final decision of a community of voters
(3,5-13).

These models focus on the self-organization resulting from a local
dynamics, which represents the mutual influence,  based on two
simple properties:

i) individuals are more likely to interact with others who already share many of their opinions;

ii) interaction increases the number of features that individuals
share.

However, a deeper analysis of the  process which leads to the final
decision in a political vote requires an improvement
of the proposed models.

This can be done in a  mathematical way by considering the
application of the game theory to political problems \cite{gg} or,
more simply, by introducing in the decision process  other
important {\bf global} effects such as, for example, the
Government policy and/or the propaganda, which are usually
neglected.

In this paper we propose a model toward a  more realistic
description of the opinion dynamics underlying the choice of the
voters. The analysis is restricted to  a bipolar scheme with a
possible third political area, but it can be generalized to the
case of many political parties. This initial simplification helps
to clarify the model and, anyway, it applies to many European and
non European countries.

The model is based on the following points to be formalized and discussed in details later on:

1) There are initially three different groups of voters: the right
Coalition (RC), the left Coalition (LC) and a central group (CG).
The voters in the RC or LC do not change their opinion: they
represent the cores of the political bipolar scheme.

2) The individuals of the CG interact with each other and with the
individuals of the RC and of the LC. At the end of the dynamical
process there are three groups: majority, minority and the others.

3) The final decision of the single CG voter is based on his/her
opinions on the arguments that he/she considers more relevant.

4) The mutual local influence among voters is modified by the
global degree of satisfaction/dissatisfaction with respect to the
Government.

The scheme (1-4) is, of course, an extremely simplified version of
the real processes. For example the possibility that each voter
(LC, RC, CG) decides not to vote (abstentionism) is not taken into
account in the dynamics and this implies that the abstentionism is
proportionally distributed among the different groups.

However the model has many steps of analysis,it  requires the
application of general  techniques of the complex systems and, as
we shall see, it is able to give some interesting indications on
the political dynamics.

The paper is organized as follows: the cognitive political model
and the social-political interaction are described in Sec. 1; in
Sec. 2 we set the criterion of  political decision; in  Sec. 3 the
approximations introduced in the numerical simulations are
discussed; the phenomenological analysis of the political vote in
Italy and Germany and the prediction for the next Italian vote are
carried out in Sec. 4; Sec. 5 is devoted to the conclusions and
outlooks.

\section{1. THE COGNITIVE POLITICAL MODEL}

\subsection{ Model general structure}

In simulating the opinion dynamics toward a final political vote one has to generalize
the available  models of social interaction.

In the Axelrod model \cite{axel,cast} of the cultural evolution
the interaction is limited to an imitation process in which the
agents adapt cultural traits stochastically from each other with a
bias toward similar agents. The interesting final result of the
evolution is the diversity: there are different cultural domains.

In the bounded confidence model \cite{def,heg,santo1}, more
oriented to the analysis of voting patterns, each voter possesses
a single real valued opinion. When two voters interact, they
average their opinions only if the opinion difference is within an
external pre-fixed threshold otherwise there is no interaction.
Also in this case, as in the previous models, the system breaks up
into distinct opinion clusters.

However, in the analysis of the political opinion dynamics there
is, ab initio,  a community with a non-negligible heterogeneity
and the selfish individual  convictions play a crucial role not
only in the interaction among individuals but also in the degree
of influence of global effects (governments, mass media,social
shocks \cite{santo2,stauf2})
 on the single voter.

In particular,in a bipolar political system, the number of votes
of the two most important (left and right) coalitions depends upon
an almost constant  core of voters, who do not  change their
opinion, and on the individuals with less strong political
convictions (the CG) who decide on the basis of a personal
political analysis.

More precisely, in the political decision process each individual
is in front of many questions of social importance (the contexts)
and he/she has to evaluate the possible alternative choices.

This analysis, based on a personal mental representation of the
validity of the different alternatives, evolves according to the
interaction with the other members of the community and according
to the global influence.

As a result of this process, a restricted number of more relevant
concepts emerges: in the individual decision making mechanism
there is a simplification with respect to the social-political
complexity and a ``dimensional reduction'' to the most relevant
aspects.

The previous considerations can be formalized by following the
interesting cognitive model proposed in ref. \cite{gabor} that,
with some peculiar modifications (see later), is a good starting
point to investigate the opinion dynamics toward a realistic
simulation of the final political vote.

In ref. \cite{gabor} one considers $I$ agents who divide the world
into a number $X$ of {\em contexts} in which they evaluate
alternative possibilities (scenarios) according to their personal
opinions. The alternative scenarios are characterized by their
objective attributes $\bm{a}=\{a_1,\ldots,a_D\}$. A crucial
assumption of the model is that the Agent $i$'s {\em theoretical
payoff} from the realization of a possible scenario $\bm{a}$ in
context $x$ is posited to be a linear function
\begin{equation}
  \pi_i^{(x)}(\ba) = \bomega_i^{(x)}\cdot\ba
  \label{objutil}
\end{equation}
where $\bomega_i^{(x)}$ is the agent's {\em context vector},
reflecting actual circumstances of the context and the agent's
personal opinions. For each agent there are $X$
context vectors each of dimension $D$, which are assumed fixed in
the model.
\newline
\ An agent does not know his/her context vectors explicitly as
this would require a detailed understanding of the effect of all
attributes on his payoffs. However, by collecting experience on
choices he/she has made previously, he/she learns to approximate
the payoffs using an appropriate \emph{mental representation}. The
mental representation is built around the world's $K$ ``most
important degrees of freedom'', constituting the agent's Concepts.
One assumes again that the {\em approximate payoff} that the agent
``computes'' directly is linear
\begin{eqnarray}
  \tilde{\pi}_i^{(x)}(\ba) = \tilde{\bomega}_i^{(x)}\cdot\ba
  \label{approxutil}
\end{eqnarray}
with
\begin{eqnarray}
  \tilde{\bomega}_i^{(x)} =
  \sum_{\mu=1}^K v_{i\mu}^{(x)} \bgamma_{i\mu}
  \label{approxutil2}
\end{eqnarray}
the agent's {\em approximate context vector}.
\
\newline

By eq.(3), $\tilde{\bomega}_i^{(x)}$ is decomposed using \emph{mental
weights} $v_{i\mu}^{(x)}$ in a reduced subspace of dimension $K$ and
 a number $K$ of concept vectors,
$\{\bgamma_{i\mu}\}_{\mu=1}^K$, assumed normalized
$|\bgamma_{i\mu}|=1$, which the agent uses to evaluate
alternatives.

Due to the reduction
of dimensionality, $K<D$, the approximate payoff
$\tilde{\pi}_i^{(x)}$ deviates from the theoretical payoff
$\pi_i^{(x)}$. The agents' goal is to find the best possible set
of Concepts and mental weights which minimizes the error of the
mental representation under the constraint that only $K$ Concepts
can be used. The natural measure of agent $i$'s {\em
representation error} is the variance\begin{equation}
    E_i = \sum_{x=1}^X \left\langle (\pi_i^{(x)}-\tilde{\pi}_i^{(x)})^2 \right\rangle_x
\end{equation}
where $\langle .\rangle_x$ is the average over alternatives in
context $x$, that can be written as
\begin{eqnarray}
   E_i = \sum_{x=1}^X |\bomega_i^{(x)}|^2-U_i
   \label{EiL}
\end{eqnarray}
where  the agent's utility $U_i$ is given by

\begin{eqnarray}
   U_i  &=&
        \sum_{\mu=1}^K \bgamma_{i\mu}\cdot \bW_{i}\ \cdot \bgamma_{i}^\mu
   \label{Utility}
\end{eqnarray}
and $W_i$ is a positive semi-definite, $D\times D$ dimensional
matrix, called the {\em \textbf{world matrix}},
\begin{equation}\label{worldmatrix}
\bW_i= \sum_{x=1}^X \bomega_i^{(x)}\circ\bomega_i^{(x)},
\end{equation}
which encompasses all information about agent $i$'s
relationship to the world.

The criterion for the political decision, that will be discussed
in sec. 2, is based on  the minimization of  $E_i$ which is
equivalent to maximize $U_i$. This  is the well-known
\textbf{Principal Component Analysis} (PCA) problem. According to
this,  the optimal concept vectors are provided by the $K$ most
significant (largest eigenvalues) eigenvectors of $\bW_i$ in the
considered basis of the D dimensional space (see below). Thus to
achieve the best possible mental representation the agent should
choose his/her concept vectors according to the eigenvectors of
his/her world matrix in the order of their significance.

Two important remarks are now in order:

a) In the application of the  model to the political dynamics the
previous considerations apply only to the agents in the CG (see
the next subsections). The agents in the RC and LC are assumed to
have fixed orthogonal concept vectors ,called  $r_\mu$ and $l_\mu$
respectively, with scalar products
 $r_\mu  r_\nu = \delta_{\mu \nu}$, $l_\mu l_\nu = \delta_{\mu \nu}$,  $r_\mu  l_\mu = 0 $.
These vectors form the basis of two fixed orthogonal subspaces.

The orthogonality of the subspaces associated with the LC and the
RC is also an approximation because in the real dynamics the
concept vectors of the LC and RC are not always completely
orthogonal (bipartisan choices). With this approximation one
assumes that the bipartisan choices are irrelevant for the final
political decision.

b) The identification, sorting and truncation of the degrees of
freedom in the model is closely analogous to what occurs in
White's Density Matrix Renormalization Group method
(DMRG)\cite{White1992,whitedmrg}). In the DMRG the optimally
renormalized degrees of freedom turn out to be the $K$ most
significant eigenvectors of the reduced density matrix of the
quantum subsystem embedded in the environment with which it
interacts.

\subsection{The socio-political interaction}

As discussed in the introduction, the opinion dynamics is due to
local interactions among voters  and to global political effects.
The local and global interactions, related with the previous
cognitive  model, will be now separately analyzed.
\
\newline
\newline
{\bf a) The local interaction}

The representation error is minimal if the agent
learns to approximate his/her world matrix in the $K$ dimensional
subspace spanned by the most significant eigenvectors of his/her world
matrix. The final vote decision is due to the social-political network
and, in our bipolar scheme,it is useful to cast context vectors into two
basic categories \cite{gabor}:
\newline
\newline
1)  context vectors which  only depend on a single agent,
 with a world matrix $W^0_i$
\newline
\newline
2) context vectors for agent $i$ which depend on the
interaction with  at least one other agent $j$.
\newline
\newline
For simplicity only pair interactions will be considered and following ref. \cite{gabor}
the world matrix  to be used in the utility function is written as

\begin{equation}
    \bW_i = \bW^0_i + \sum_{j\in {\cal CG}} c_{ij}\, \bS_j +  \sum_{j\in {\cal LC}} l_{ij}\, \bS_j +
 \sum_{j\in {\cal RC}} r_{ij}\, \bS_j,
    \label{culture1}
\end{equation}
where the total agents number is ${\cal N}$, the agent $i$ is in
the CG (see later), the parameters $c_{ij}$, $l_{ij}$, $r_{ij}$
measure the relative strength  of socio-political interactions
with agent $j$ in the different groups (CG, LC, RC) and $S_j$ is
given by
\begin{equation}
    \bS_j=\sum_{\mu=1}^K \bgamma_{j\mu}\circ\bgamma_{j}^\mu,
\end{equation}
i.e. by  $S_j$ there is in $W_i$ an overlap of the concept vectors of the agents
$i$ and $j$.

It is important to stress that  the previous interactions are among the agents $i$ of the CG with
the other agents in the same group, in the LC and in the RC.
Indeed, in our scheme the agents in the LC or RC do not change their opinion and their world matrices are
fixed to  constant $W^0_i$.  Moreover
the world matrix of the agents in the LC is orthogonal to the world matrix for the RC agents.

\
\newline

{\bf b) The global effects}

The  opinion making process in a national political vote  depends
only partially on the local interaction and  is strongly
influenced by other important elements such as the decisions of
the Government and the mass media role. In order to include these
effects in the dynamical process, let us introduce a set of
indices, $\delta_i$, to describe the satisfaction of the agent $i$
with respect to the global political decisions. Again the voters
in the RC and LC do not change opinion despite their
satisfaction/dissatisfaction and then the satisfaction indices, $0
< \delta_i < 1$, related with  the rating of the global events,
are relevant only for the voters in the CG.

In the previous subsection, the local influence has been
introduced by the political interaction matrices $c_{i,j}$,
$l_{i,j}$, $r_{i,j}$.
 In the political bipolar scheme, a simple way to include  the
satisfaction indices, $\delta_i$, in the dynamics is to consider
that a voter who has a positive perception of the Government
actions  increases the strength of its interaction with the
coalition  which governs while a dissatisfied voter tends to
interact more with the opposition. Therefore the  interaction
matrices  of an agent $i$ of the CG with the agents $j$ of the LC
or of the RC will  depend on  the satisfaction index in such a way
to increase the interaction with the majority and decrease the
interaction with the minority or viceversa. The interaction among
voters in the CG remains unchanged.

In section 3 the $l_{i,j}$, $r_{i,j}$ dependence on $\delta_i$
will be clarified.

\section{2. The decision process}

The most important elements which  determine
the  agent representation  of the social-political system have been specified in the previous sections.
Following ref. \cite{gabor}, one can assume  that the  system evolves according to a
``gradient adjustment  dynamics'' obtained by the time evolution of the concept vectors given by
\begin{equation}
\frac{\delta \gamma_{i\mu}}{\delta t} \propto \frac{\partial U_i} {\partial \gamma_{i\mu}}.
\end{equation}
Of course, only the  CG agents have a dynamical evolution
 because the agents in the LC and in the  RC
have fixed world matrices , $\bW_i=\bW^0_i$, and the corresponding concept vectors span orthogonal subspaces.
For the agents in the CG it is reasonable to assume that
without interaction they choose random concept subspaces, i.e. complete disorder,
and this implies that their  matrices  $\bW^0_i$ have a  Wishart distribution (see
\cite{Wishart1928,Edelman1989}).
After the evolution according to the best response dynamics, at the equilibrium \cite{gabor},
each agent of the CG  decides his/her vote
and one needs a well defined criterion to understand if his/her final ``position''
is closer to the LC or the RC or is too far from both.
The most simple idea is a comparison of the final concept vectors of each agent in the CG with the
concept vectors of the LC and of the RC and then to define a ``political distance'' by the scalar
products of the previous concept vectors.
However, as clarified in ref.\cite{gabor}, the choice of the concept vectors
is not unique and only the subspace they span is relevant for the dynamical process.
Then the criterion has to be related with the subspace spanned by CG agent concept vectors at the equilibrium with respect
to the LC subspace (i.e. spanned by the concept vectors  of the LC agent) or to the RC subspace and  it is natural
to consider the ``angle'', $\theta$,  between two subspaces as the agent-agent distance \cite{gabor2}.

The procedure is the following one.
Firstly one considers the, $K$,  $\gamma_i$ concept vectors of agent $i$ as columns
of a new matrix $\Omega_i$ ($D$ x $K$ dimensional matrix).
Similarly we construct the matrix $\Omega_j$ of any agent of LC and of the RC (fixed by definition).
According to refs. \cite{bjorck,wedin}, we calculate the overlap between
$\Omega_i$ and $\Omega_j$ in the following way:

\begin{eqnarray}
\Omega_i^{'}=\Omega_i-\Omega_j (\Omega_j^T \Omega_i)
\end{eqnarray}

Hence, we calculate the angle $\theta$ (in $[0,\pi/2]$) as

\begin{eqnarray}
\theta_{i,L}=\arcsin (\min[1,\|\Omega_i^{'}\|_{\infty}])
\end{eqnarray}
and
\begin{eqnarray}
\theta_{i,R}=\arcsin (\min[0,\|\Omega_i^{'}\|_{\infty}])
\end{eqnarray}
where $\|\Omega_i^{'}\|_{\infty}$ is the biggest singular value of
$\Omega_i^{'}$ according to the Singular Value Decomposition
(SVD). With this definition the distance is between $0$
(politically equivalent) and $\pi/2$ (politically  orthogonal).
Finally the agent $i$ will vote for the RC or the LC, according to
the fact that the smallest ``angle'' is $\theta_{i,R}$ or
$\theta_{i,L}$, respectively.

However, in order to have a more realistic model of the decision
process, one has to take into account another important political
aspect: the bipolar systems are, indeed, not perfectly bipolar.
There is a non negligible part of the  individuals in the CG that
at the end of the dynamical evolution is still ``too far'' from
the political ideas of the L and the R coalition and decide to
vote for other possible groups (OG).

Of course one can neglect this point, which, however, is a crucial ingredient of the
political dynamics.

Our definition of ``political distance'' between two agents is related to the angle $0 <\theta < \pi/2$
of the corresponding subspaces spanned by the concept vectors. When $\theta=\pi/2$ two subspaces are orthogonal
and then an agent $i$ of the CG
will be considered voting for the OG if {\bf both} the angles $\theta_{i,R}$ and $\theta_{i,L}$ are
in the range between a fixed value  $\epsilon$ and $\pi/2$. An increase of $\epsilon$ decreases the number of agents
in the CG who vote for the third group: $\epsilon$ represents the ``mobility'' of the CG toward the political coalitions.

\section{3. The approximations}

The  equilibrium and dynamic properties depend on the
interaction matrices  $c_{ij}$, $l_{ij}$, $r_{ij}$.
As a first approximation, one considers for the {\bf local} interaction
a  mean-field political network with $c_{ij} = l_{ij} = r_{ij} = h $.
This implies that, without global effects, the  final voting choice is directly correlated with the
number of agents of the LC and of the RC: a larger number of agents in the LC (RC)
will automatically determine that a larger number of CG agents will we ``closer'' to the LC (RC).

Moreover, the global effects change the interaction strength of
the agent $i$ in the CG with the voters of the L and R coalitions,
by the satisfaction index $\delta_i$ (the interaction within the
CG is unmodified).

For simplicity, let us assume that $\delta_i = \delta$ is
independent on the agent $i$. By definition, it has opposite
effect on the interactions with the LC and the RC. Then the
starting approximation for the world matrix $W_i$ of the agent $i$
in the CG, which includes the local and the global interactions,
can be written as
\begin{equation}
    \bW_i = \bW^0_i + \sum_{j\in {\cal CG}_i} h \bS_j +  \sum_{j\in {\cal LC}_i} h(1 + \delta) \bS_j +
 \sum_{j\in {\cal RC}_i} h(1 - \delta) \bS_j,
    \label{culture1}
\end{equation}

Where $\delta$ has been conventionally assumed positive if the interaction with the LC increases.
The parameter $\delta$ drives the opinion dynamics and determines the majority. A more detailed
analysis of its meaning and effects is carried out in section 4.

Another assumption concerns the initial core distributions of the
L and the R coalitions. In fact, it is not difficult to take into
account a numerical difference between the  core voters of the two
coalitions,  but this would introduce another parameter in the
model and, in the present work,  we are mainly interested in
analyzing the global effects on the vote decision. Therefore, in
this first version of the model, we shall consider a symmetrical
initial distribution for the core voters of the L and R
coalitions. Moreover, abstentionism is not of dynamical origin and
then, in the numerical simulation, is  proportionally divided
according to the group initial distributions.

The initial core voters of the L and  R coalitions in our
simulations are assumed equal to the  $41 \%$  of valid votes
while the CG has only the $18 \%$.

These numbers are not far from the real political situation: we
verified that the lowest level for the L and R coalitions (assumed
as the  cores) in Italian national votes, from 1996 to 2004, is
close to the $41 \%$ of the valid  votes.

This implies that the winning coalition is determined by the
decision of a relatively small number of individuals of the CG.

\section{4. Vote analysis}

Before analyzing the data of the national vote in Italy and in Germany, let us note that  in the  model there
are essentially two parameters, $\delta$ and $\epsilon$, because, in the data fitting,
a change in the parameter $h$ gives a rescaling of the previous  ones.

For the simulations of the dynamical evolution one considers the
initial configuration previously discussed, with D=10, K=4 and
10.000 agents. The results are averaged on 200 samples.

\subsection{Italian elections}

Table \ref{tab1} reports the results of the Italian national vote
from 1994 to 2004 for the different groups, the difference (in
percent), $\Delta_{LR}$, in the final vote between the L and R
coalitions and the fitted values of $\delta$ and $\epsilon$ which
reproduce the data (starting with the initial configurations
previously discussed).

\begin{table}[ht]
\caption{\label{tab1}The results of the Italian national vote from 1994 to 2004 for the three different groups (RC, LC, OG) and the difference ($\Delta_{LR}$) (in percent). We also report the respective values of $\delta$ and $\epsilon$ which allow us to reproduce the data.
The L coalition includes the following parties: Democratici di Sinistra,
Margherita, Rifondazione Comunista, Verdi, Udeur, SDI, Italia dei valori, Comunisti Italiani.
The R coalition contains Forza Italia, Alleanza Nazionale, Lega Nord, UDC, PRI.
The other parties are in the third group.}
\begin{ruledtabular}
\begin{tabular}{ccccccc}
Year&RC&LC&OG& $\Delta_{LR}$ & $\delta$ & $\epsilon$\\
\hline
1994 & 50.4\% & 45.3 \% & 4.3 \%& -5.1 \% & -0.555 & 1.465\\
1996 & 42.2\% & 43.3 \% & 14.5 \%& 1.1 \% & 0.134 &1.356\\
1999 & 43.8\% & 42.2 \% & 14.0 \%& -1.6\% & -0.195&1.361\\
2001 & 49.5\% & 46.3 \% & 4.2 \%& -3.2 \% & -0.345 &1.468\\
2004 & 46.6\% & 45.5 \% & 7.9 \%& -1.1 \% & -0.121&1.425\\
\end{tabular}
\end{ruledtabular}
\end{table}

The fit shows that when $\epsilon$ increases, the voters of the OG decrease and the
final vote is  politically more polarized: the difference between
the two leading coalitions increases.

Moreover, when the (dis)satisfaction index $\delta$ is small also
$\epsilon$ is small, i.e. the agents of the CG prefer to vote for
a third group. In other words, when the  consequences of  the
government choices are considered ``neutral'' by the CG agents,
the latter tend not to vote for a more political  coalition.

On the other hand, when the (dis)satisfaction increases, they feel
that a stronger political option is more useful and vote for the L
or the R coalition. There is anyway a ``physiological'' threshold
of about 4 $\%$ of voters who choose the OG.

The picture becomes more clear if one looks carefully the table
and considers the evolution of the different Italian governments
from 1994 to 2004.

In 1994 the RC won the election with a strong dissatisfaction
toward the previous government ($\delta= -0.55$ and $\epsilon=
1.465$).

In 1996, due to the internal problem of the majority, a party left
the RC and created an independent pole. This increased the OG and
gave the final victory to the LC (by a moderate dissatisfaction
$\delta=0.134$ and $\epsilon=1.36$).

In 1999, the first negative signals for the LC government appeared
as a small dissatisfaction ($\delta=-0.195$) and a
 larger vote to the  OG ($\epsilon=1.36$) in the results
of the election for the  European Parliament.

In 2001 the national vote was  polarized against the LC government ($\delta=-0.345$ and $\epsilon=1.47$)
and the RC won the election.

The European  vote in 2004 is, in turn, a negative indication for
the RC government because there is a strong reduction of $\delta$
($\delta= -0.12$ with respect to the previous value $\delta=
-0.345$) and a clear signal of an emerging larger third group with
respect to the previous election in  2001.

Hence the general behaviour of the  agents in the CG seems to
develop according the following steps: i) an initial political
choice; ii) a dissatisfaction that leads to vote for the OG in the
mean term elections; iii)a polarization against the Government in
the following political election.

The previous considerations are  probably  obvious
for an ``educated'' political observer and yet it is quite  interesting that
they are reproduced by a mathematical model.
 Moreover one can play a risky game: a prediction for the next Italian vote in
April 2005 (see later).

\subsection{ German elections}

Let us now consider the national  vote in Germany from 1987 to 2002.

Since we also want  to have indications about
the different political behaviour of the Italian and German voters, we shall start with the previous initial
distribution among the RC, the LC and the
CG which, anyway, is not far from the minimum results obtained by the R and L coalitions
($41.3 \%$ and $39.7 \%$).

Table \ref{tab2}, as table \ref{tab1} before, reports the results
of the German national vote for the different groups and  the
fitted values of $\delta$ and $\epsilon$.

\begin{table}[ht]
\caption{As in tab. \ref{tab1} the results of the German national
vote from 1987 to 2002 for the three different groups (RC, LC, OG)
and the difference ($\Delta_{LR}$) (in percent). We also report
the respective values of $\delta$ and $\epsilon$ which allow us to
reproduce the data. The L coalition is given by SPD, Green party,
PDS and the R coalition includes CDU, CSU and FDP. The other
parties are in the third group.\label{tab2}}
\begin{ruledtabular}
\begin{tabular}{ccccccc}
Year&RC&LC&OG& $\Delta_{LR}$ & $\delta$ & $\epsilon$\\
\hline
1987 & 53.4\% & 45.3 \% & 1.3 \%& -8.1 \% & -0.944 & 1.513\\
1990 & 54.8\% & 39.7 \% & 5.5 \%& -15.1 \% & -2.215 &1.410\\
1994 & 48.3\% & 48.1 \% & 3.6 \%& -0.2\% & -0.013&1.477\\
1998 & 41.3\% & 52.7 \% & 6.0 \%& 11.4 \% & 1.440 &1.430\\
2002 & 48.5\% & 47.7 \% & 3.8 \%& -0.8 \% & -0.080&1.475\\
\end{tabular}
\end{ruledtabular}
\end{table}

There is a clear relation  between $\epsilon$ and $\delta$ which
shows that  German voters have different attitude with respect to
the Italian ones. Indeed, in this case, when $\delta$ increases
$\epsilon$ decreases, i.e. the third group is larger. In other
words, a bad government rating leads to vote for other groups
rather than for the political coalitions. This is probably due to
the fact that in Germany the other groups are essentially
politically oriented while in Italy they tend to be  more
independent ones.

A deeper analysis requires, again, an understanding of the
dynamics of abstentionism.

\subsection{The meaning of the variation of $\delta$}

The (dis)satisfaction index $\delta$ describes the rating of the
government decisions as perceived by the agents and the parameter
$\epsilon$ is a measure of the mobility of the CG.

With a positive/negative  $\delta$ the majority will be of the
LC/RC coalition. However to have the majority of votes does not
automatically mean that the corresponding coalition wins the
election:  a coalition which obtained  a  $60 \%$  majority in the
last  national vote and obtains only the $55 \%$ in the next one
has still the majority (and then $\delta$ has the same sign) but
with less consensus and, in this sense, does not win the election.

Therefore $\delta$ is effectively the index of the  strength of
the interaction of the CG with the L an R coalitions which
determines the majority. Its variation from an election to the
next one, $\Delta \delta$, is a more reliable ``measure'' of the
evolution of the rating of the government and of the success of
the coalitions.

\begin{figure}[th]
\includegraphics[width=0.46\textwidth]{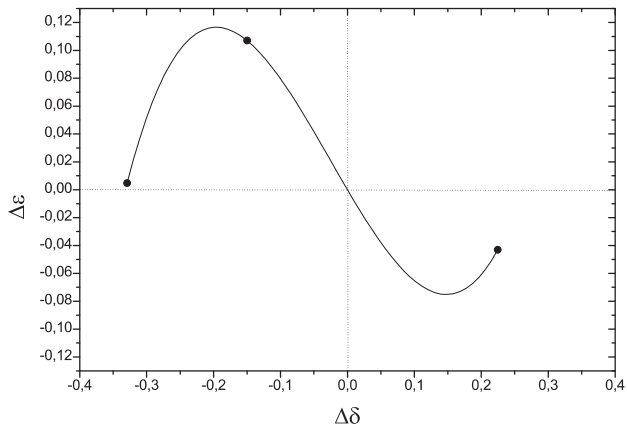}
\caption{\label{fig1}Correlation between $\Delta \delta$ and $\Delta \epsilon$ for the Italian case.}
\end{figure}

The values of $\Delta \delta$ are obtained by table
\ref{tab1}: $\Delta \delta (99-96)= -0.33$, $\Delta \delta
(01-99)=-0.15$, $\Delta \delta (04-01) = 0.22$ (the elections
before these dates are not considered because the victory of the
LC in 1996 was due to a breaking of the RC and the  government
before the 94 election was a technical government rather than a
political one) and they are correlated with the corresponding variations
of $\epsilon$, $\Delta \epsilon$, as shown in fig. \ref{fig1}.

The meaning of this correlation between  $\Delta \delta$ and
$\Delta \epsilon$ has been previously discussed and for the German
vote one has a different behaviour with respect to the Italian
case.

\subsection{ A prediction: the next vote in Italy }

It is useful to show how  the model can make some predictions and,
as an example,  let us consider
the next Italian vote in April 2005.

For the prediction one needs to know the  correlation between
$\Delta \delta$ and $\Delta \epsilon$. In fig. \ref{fig1} there
are very few points and a reliable result requires a precise
determination of the correlation. However, since it is useful, in
our opinion, to show the procedure, let us consider in the
numerical simulation the correlation obtained by fig. \ref{fig1}.

Unfortunately, the next Italian  vote is not
a national homogeneous ones but, rather, it is a regional election
extended to the large part of the country (see later).

If one considers this election  more politically homogeneous to
the vote for the  European Parliament, i.e. the crucial elements
of the political debate are similar to the previous 2004 vote,
therefore the starting point for the simulation is the election in
2004 and the results of the model as a function of $\Delta \delta
(05-04)$ is reported  in tab. \ref{tab3}. As in the previous
analyses, abstentionism is fixed to $17 \%$.

\begin{table}[th]
\caption{The prediction of the model for the next Italian vote, when the 2004 election has been considered as the starting point.\label{tab3}}
\begin{ruledtabular}
\begin{tabular}{ccccc}
$\Delta \delta$&$\Delta \epsilon$ & RC & LC & OG \\
\hline 0&0 & 46.6 \% & 45.4 \% & 8.0 \% \\
0.05& -0.038 & 44.6 \% & 43.8 \% &11.6 \%\\
0.10& -0.065 & 43.1 \% & 42.8 \% & 14.1 \%\\
0.15& -0.075& 42.4 \% & 42.6 \%&  15.0 \%\\
0.20& -0.061&  42.8 \% & 43.5 \%& 13.7 \% \\
0.25& -0.016 & 44.7 \% & 45.8 \%& 9.5 \%\\
0.30& 0.066 & 47.2 \% & 48.8 \% & 4.0 \%\\
0.35& 0.192 & 47.0 \%& 49.0 \%& 4.0 \%\\
0.40& 0.367 & 46.7 \%& 49.3 \% & 4.0 \%\\
0.45& 0.598& 46.5 \%& 49.5 \%& 4.0\%\\
0.50& 0.891 & 46.3 \%& 49.7 \%&4.0 \%\\
\end{tabular}
\end{ruledtabular}
\end{table}

On the other hand, if the next Italian vote is considered
politically more similar to a national government election, the
starting point of the analysis is the political result in 2001 and
the prediction as a function of  $\Delta \delta (05-01)$ are given
in tab. \ref{tab4}.

\begin{table}[th]
\caption{The prediction of the model for the next Italian vote, when the 2001 election has been considered as the starting point.\label{tab4}}
\begin{ruledtabular}
\begin{tabular}{ccccc}
$\Delta \delta$&$\Delta \epsilon$ & RC & LC & OG \\
\hline 0&0 & 49.7 \% & 46.3 \% & 4.0 \% \\
0.05& -0.038 & 47.7 \% & 44.8 \% &7.5 \%\\
0.10& -0.065 & 46.2 \% & 43.7 \% & 10.1 \%\\
0.15& -0.075& 45.5 \% & 43.5 \%&  11.0 \%\\
0.20& -0.061&  45.9 \% & 44.4 \%& 9.7 \% \\
0.25& -0.016 & 47.7 \% & 46.7 \%& 5.6 \%\\
0.30& 0.066 & 48.3 \% & 47.7 \% & 4.0 \%\\
0.35& 0.192 & 48.0 \%& 48.0 \%& 4.0 \%\\
0.40& 0.367 & 47.8 \%& 48.2 \% & 4.0 \%\\
0.45& 0.598& 47.6 \%& 48.4 \%& 4.0\%\\
0.50& 0.891 & 47.3 \%& 48.7 \%&4.0 \%\\
\end{tabular}
\end{ruledtabular}
\end{table}

As one can see, worsening government rating,
which implies a positive  $\Delta \delta$, can cause  today
majority to loose the election and become minority. However, this
depends on the numerical value of  $\Delta \delta$ and the
question arises if it is somehow related to the evolution of the
opinions on the Government decisions as ``detected'', for example,
by  polls.

The answer to this risky question requires a careful comparison
with the rating of the governments obtained by polls immediately
before the various elections \cite{swg}. This point is interesting
also for practical reasons (see sec. 5) and will be discussed in a
forthcoming paper \cite{cc2}.

\section{5. conclusions and outlooks}

The proposed model is a first step towards a more realistic
description of the opinion dynamics of the political vote and,
indeed, it has many key elements of the decision process. However
there are some approximations:

1) Abstentionism has not a dynamical origin.

This is an important point because the cores of the different
coalitions do not change opinion, regardless of the  value of the
satisfaction index.

However, in the real political arena the winning coalition is
determined not only by the vote decision of the agents in the CG
but also by the decision not to vote of the agents in the  core
coalition: an agent with clear political convictions but strongly
dissatisfied from  his/her government does not vote for the
opposite coalition but  prefers the abstentionism. Since the
winning coalition has only a few percent of vote more than the
minority, this effect can easily determine the final result.

2) For the same reason,the symmetric distribution of the  core of
the L and R coalitions is an approximation: a small initially
asymmetry  of the coalitions can change the final result and/or
increase the strength of the global parameter $\delta$ needed for
the victory.

3) The symmetric mean field local interaction of the agents in the CG with the LC and RC individuals.

Indeed, the opinion dynamics  starts soon after a political vote
and completes the evolution with the next vote choice. The agents
in the CG who voted for the winning coalition are, at least
immediately after the vote, predominately interested to  interact
with the majority. From this point of view, the symmetric local
interaction means that the model applies for timeframe close to
the next vote or, generally,  when the agents of the CG become
more independent respect to the previous majority. The points 1-3
require the introduction of (at least) other three parameters in
the model.

Despite  the previous approximations, that can be overcome by a more complete version,
the model is able to:

a)  describe the simplification procedure of the social-political context;

b)  introduce local and  global interaction in the dynamics;

c)  define a reliable ``political distance'' among the individuals;

d)  introduce the alternative  third choice for the voters;

e)  determine phenomelogical relations among the  parameters and some  quantities obtained by data;

f)  make some predictions;

g)  potentially combine simulations and polls.

This last point is  interesting  because  during  the political
polls,  people are more incline to answer on  more general and
less direct  political questions. For example, the answer is less
uncertain if the question is about the coalition rather than the
single political party. Now, it is not difficult to think of a set
of  undirect questions able to determine the parameters of the
model ( $\delta$, $\epsilon$ plus the others needed for points
1-3) and put on a more rigorous basis the relation among polls and
mathematical models. Finally, the  next steps include: a) a more
realistic version of the model \cite{cc2}; b) a combined effort
with poll experts; c) the application of the model to different
countries. Indeed, the correlation, if any, between $\delta$ and
$\epsilon$ certainly depends on the peculiar characteristics of
the considered nation but it should be important to verify if
there are similar political behaviors in different places.

\ \newline {\bf Acknowledgements.} The authors thank G. Fath and
D. Zappala' for useful suggestions and comments, S. Fortunato and
G. Gambarelli for interesting discussions and  R. Fonda, of the
SWG, for the help in understanding  the relation among the model
parameters and the political poll results.

\end{document}